\documentstyle[12pt]{article}
\catcode`\@=11

\@addtoreset{equation}{section}

\catcode`\@=11

\def\section{\@startsection{section}{1}{\z@}{3.5ex plus 1ex minus
   .2ex}{2.3ex plus .2ex}{\large\bf}}

\newcommand{\be}{\begin{equation}}
\newcommand{\ee}{\end{equation}}
\newcommand{\bea}{\begin{eqnarray}}
\newcommand{\eea}{\end{eqnarray}}
\newcommand{\none}{\nonumber \\}
\newcommand{\half}{{1\over2}}
\def\Idoubled#1{{\rm I\kern-.22em #1}}

\newcommand{\DA}{D_{(A)}}
\newcommand{\req}[1]{Eq.\ (\ref{#1})}
\newcommand{\reqs}[1]{Eqs.\ (\ref{#1})}
\begin{document}
\begin{titlepage}
\begin{flushright}
PACS 04.70.Dy\\
UNB Technical Report 94-03\\
UCD-94-36\\
gr-qc/9410021\\
October 1994\\
\end{flushright}
\vspace{.25in}

\begin{center}
{\large\bf Black Holes in\\[2ex] Three-Dimensional Topological Gravity}\\
\vspace{15pt}
{\it by}\\
\vspace{5pt}
S.\ Carlip \\[5pt]
{\it Department of Physics}\\
{\it University of California}\\
{\it Davis, California 95616 USA}\\
{[e-mail:  carlip@dirac.ucdavis.edu]}\\
\vspace{13pt}
J.\ Gegenberg \\[5pt]
{\it Department of Mathematics and Statistics}\\
   {\it University of New Brunswick}\\
   {\it Fredericton, New Brunswick
   \it Canada E3B 5A3}\\
{[e-mail:  lenin@math.unb.ca]}\\
\vspace{13pt}
R.\ B.\ Mann\\[5pt]
{\it Department of Applied Mathmatics and Theoretical Physics}\\
{\it Silver Street}\\
{\it Cambridge University}\\
{\it Cambridge
U.K.
CB2 9EWf}\\
and \\
{\it Department of Physics}\\
{\it University of Waterloo}\\
{\it Waterloo, Ontario Canada N2L 3G1}\\
{[e-mail:  rbm20@damtp.cambridge.ac.uk]}\\
\end{center}
\vspace{20pt}
\begin{center}
{\bf Abstract}\\
\end{center}
We investigate the black hole solution to (2+1)-dimensional
gravity coupled to topological matter, with a vanishing cosmological
constant.  We calculate the total energy, angular momentum and entropy
of the black hole in this model and compare with results obtained in
Einstein gravity.  We find that the theory with topological matter
reverses the identification of energy and angular momentum with the
parameters in the metric, compared with general relativity, and that
the entropy is determined by the circumference of the inner rather
than the outer horizon.  We speculate that this results from the
contribution of the topological matter fields to the conserved
currents.  We also briefly discuss two new possible (2+1)-dimensional
black holes.
\end{titlepage}

\section{The (2+1)-Dimensional Black Hole}

It is a peculiar feature of general relativity in 2+1 dimensions that
any solution of the Einstein field equations
\begin{equation}
G_{\mu\nu} = 8\pi G T_{\mu\nu} - \Lambda g_{\mu\nu},
\label{eq1}
\end{equation}
with vanishing stress-energy tensor has constant curvature.  Despite
this limitation, Ba\~nados {\it et al.} \cite{btz} have made
the interesting observation that when $\Lambda=-1/\ell^2<0$, the field
equations have a black hole solution, characterized by the metric
\be
ds^2 = -N^2 dt^2 + N^{-2}dr^2+r^2(N^{\phi}dt+ d\phi)^2
,\label{eq5}
\ee
$$
\quad -\infty<t<\infty\,,\quad 0<r<\infty\,,\quad
0\leq \phi\leq 2\pi,
$$
with lapse and angular shift functions
\be
N^2(r) =-M+{r^2\over \ell^2}+{J^2\over 4r^2} ,\qquad
N^\phi(r)=-{J\over 2r^2} .
\label{eq7}
\ee
As a space of constant curvature, this geometry can be obtained directly
from anti-de Sitter space by means of appropriate identifications, as
discussed in Ref.\ \cite{bhtz}.  When  $M>0$ and $|J|\leq M\ell\,$, the
solution has an outer event horizon at $r=r_+$, where
\be
r_+^2={M\ell^2\over 2}\left \{ 1 +
\left [ 1 -\left({J\over M\ell}\right )^2\right ]^{1/2}\right \} ,
\label{eq8}
\ee
and an inner horizon at $r_- = {J\ell/ 2r_+}$.

The parameters $J$ and $M$ have been shown to be the quasilocal
angular momentum and mass of the black hole \cite{bcm}; alternatively they
can be expressed in terms of Casimir invariants in a gauge-theoretic
formulation of $(2+1)$-dimensional gravity \cite{clm}. The parameter $M$
can also be expressed in terms of the initial energy density of a disk of
collapsing dust in AdS space \cite{Ross}.

For later reference, it will be useful to display the first order formulation
of Einstein gravity and the first order form of the black hole solution.
We suppose that $M^3$ is a smooth orientable 3-manifold whose cotangent
bundle has SO(2,1) as its structure group.  The fibers of $T^*M^3$
are three-dimensional vector spaces which come equipped with a ``natural''
metric $\eta_{ab}$ and volume element $\epsilon_{abc}$.  A smooth frame
field, or triad, on $M^3$ is a set of three independent one-form fields
$E^a$, and a spin connection $\omega_a$ on $M^3$ is an SO(2,1) connection.
In terms of these fields, the Einstein action takes the simple form
\cite{witten}
\be
I= \int {\bf L} = -\half\int_{M^3}\left( E^a\wedge R_a[\omega]
- \frac{\Lambda}{3}\epsilon_{abc}E^a\wedge E^b\wedge E^c\right),
\label{eq9}
\ee
where the curvature $R_a[\omega]$ is
\be
R_a[\omega] = d\omega_a+{1\over 4}\epsilon_{abc}\,\omega^b\wedge\omega^c .
\label{curv}
\ee
We use the convention $\epsilon^{012}=+1$, and our units are
such that $8\pi G=1$.  Variation of $I$ with respect
to $\omega_a$ yields the condition
\be
D_{\omega}E^a = dE^a+ \half\epsilon^{abc}\,\omega_b\wedge E_c=0,
\label{eq10}
\ee
which is the usual relationship between $E^a$ and the connection (that
is, the condition that $\omega^a$ be torsion-free).  Variation with respect
to the triad yields
\be
R_a = \Lambda\epsilon_{abc}E^b\wedge E^c,
\label{eq11}
\ee
which is equivalent to \req{eq1} upon insertion of \req{eq10}.

The black hole (\ref{eq5}) can be now described by the spacetime triad
\bea
E^0& =&\sqrt {\nu^2(r)-1}
\left({r_+ \over\ell}dt - r_- d\phi\right),\none
E^1&=&{\ell\over \nu}d(\sqrt{\nu^2(r)-1}),\none
E^2&=&\nu(r)\left (r_+ d\phi - {r_-\over\ell}dt \right)
,\label{btztriad}
\eea
and the compatible spin connection
\bea
\omega^0&=&-2\sqrt{\nu^2(r)-1}
\left({r_+\over\ell}d\phi - {r_-\over \ell^2}dt \right),\none
\omega^1&=&0,\none
\omega^2&=&-2\nu (r)\left({r_+\over \ell^2}dt -
{r_-\over \ell} d\phi \right) ,
\label{btzconn}
\eea
where $\nu^2(r) := \frac{r^2 - r^2_-}{r^2_+ - r^2_-}$.

\bigskip
\section{BCEA Gravity}

There is another three-dimensional theory of gravity that admits the
aysmptotically AdS black hole as a solution.  This theory, developed by
two of the present authors \cite{cargeg}, minimally couples topological
matter to Einstein gravity in 2+1 dimensions, in effect replacing the
cosmological constant by a simple set of matter fields.  We shall call
this model ``BCEA theory.''

We start with a triad $E^a$ and a spin connection $A^a$ as in the last
section, but without imposing the torsion-free condition (\ref{eq10}).
Now let $B^a$ and $C^a$ be two additional one-form ``matter'' fields.  The
action for BCEA theory is
\be
I= \int {\bf L} =
-\half\int_{M^3}\left(E^a\wedge R_a[A]+B^a\wedge \DA C_a\right) ,
\label{bceaaction}
\ee
where $\DA$ is the covariant derivative with respect to the
connection $A^a$ and the curvature $R_a[A]$ is given by
(\ref{curv}).

The stationary points of $I$ are determined by the field equations
\bea
R_a[A]&=0, \nonumber\\
\DA B^a&=0, \nonumber\\
\DA C_a&=0, \nonumber\\
\DA E^a+\half \epsilon^{abc}B_b\wedge C_c&=0.
\label{eq:7d}
\eea
Because of the term in $B_b\wedge C_c$ in the last equation of motion, the
triad $E^a$ is not, in general, compatible with the spin connection $A_a$.
Nevertheless, the equations of motion above determine a Lorentzian geometry
on $TM^3$: if we define a one-form field $Q_a$ by the requirement
\be
\epsilon^{abc}\left(Q_b\wedge E_c -B_b\wedge C_c\right)=0 ,
\label{aa0}
\ee
then the equation of motion for the $E^a$ can be written as
\be
dE^a+\half\epsilon^{abc}\omega_b\wedge E_c=0, \qquad
\omega_a:=A_a+Q_a .
\label{aa}
\ee

\req{aa} may be recognized as the condition that the frame field $E^a$
be compatible with the (nonflat) spin connection $\omega_a$.
We may thus interpret BCEA theory as a model of (2+1)-dimensional gravity
with a triad $E^a$ and a connection $\omega_a$ coupled to matter fields
$B^a$ and $C^a$.  Alternatively, the model may be viewed as a
``teleparallel'' theory of gravity, with a triad $E^a$ and a flat---but
not torsion-free---connection $A^a$, again coupled to matter fields
$B^a$ and $C^a$.  In either case, the geometry is determined by the
metric $g_{\mu\nu} = \eta_{ab}E^a{}_\mu E^b{}_\nu$.

The action functional $I$ is invariant under a twelve-parameter group
whose infinitesimal generators are \cite{cargeg}
\bea
\delta B^a&=&\DA \rho^a+\half\epsilon^{abc}B_b\tau_c,\nonumber\\
\delta C^a&=&\DA\lambda^a+\half\epsilon^{abc}C_b\tau_c,\nonumber\\
\delta E^a&=&\DA\xi^a+\half\epsilon^{abc}\left(E_b\tau_c
 +B_b\lambda_c+C_b\rho_c\right),\nonumber\\
\delta A^a&=&\DA\tau^a .
\label{eq:11}
\eea
This group may be recognized as I(ISO(2,1)), where the notation IG denotes
the semi-direct product of the Lie group G with its own Lie algebra
${\cal L}_G$.  Like the action for ordinary Einstein gravity in three
dimensions \cite{witten}, the BCEA action can be obtained from a
Chern-Simons functional, now for the gauge group I(ISO(2,1)).

\section {The Black Hole Solution in BCEA Theory}

We shall now demonstrate that the asymptotically anti-de Sitter black hole of
section 1 is also a solution of BCEA theory.  The computation is simplest
in the gauge $A^a=0$, for which the field equations for $B$ and $C$
reduce to the condition that the one-forms $B^a$ and $C^a$ be closed.  It
is then easy to show that
\bea
B^0&=& -r_+d\phi+{r_-\over \ell}dt,\nonumber\\
B^1&=&-\ell d(\nu+\sqrt{\nu^2-1}), \nonumber\\
B^2&=&{r_+\over \ell}dt -r_-d\phi
,\label{Beqn}
\eea
and
\bea
C^0&=&-{1\over \ell}B^0 ,\nonumber\\
C^1&=&d(\nu-\sqrt{\nu^2-1}), \nonumber\\
C^2&=&{1\over \ell}B^2
,\label{Ceqn}
\eea
together with the black hole triad of equation (\ref{btztriad}) and the
connection $A^a=0$, satisfy the BCEA equations of motion.

As discussed above, the triad $E^a$ determines the spacetime geometry, so
the solution given by \reqs{btztriad}, (\ref{Beqn}), and (\ref{Ceqn}) still
has an interpretation as a black hole.  To obtain more information, we can
investigate the conserved quantities, or Noether currents, associated
with the symmetries of the solution.  For the conventional black hole, for
example, the charge and mass are the conserved charges associated with
spatial translations and rotations at infinity, and the entropy is the
charge associated with an appropriate Killing vector at the horizon.  It
is instructive to understand the analogous statements for the BCEA solution.

\bigskip
\section{Noether Currents}
We shall follow the methods of Wald, which we briefly summarize for the
case of a three-dimensional spacetime (see \cite{wald} for details).

\medskip\noindent
{\bf A.  Noether Charges a la Wald}
\par
Let ${\bf L}[\phi]$ be a Lagrangian three-form (in three-dimensional
spacetime), where $\phi$ represents an arbitrary set of fields.  Under
a variation $\delta\phi$,
\be
\delta {\bf L} = \Xi\delta\phi + d{\bf\Theta}[\phi,\delta\phi] ,
\label{sc1}
\ee
where the field equations are $\Xi = 0$ and ${\bf\Theta}$ is a
boundary term, constructed locally from $\phi$ and $\delta\phi$, that
determines the symplectic structure of the theory.  For the action to
be invariant under a symmetry transformation
\be
\phi\rightarrow\phi + \delta_g\phi ,\qquad g\in G ,
\label{sc2}
\ee
it is clearly necessary that
\be
\delta_g {\bf L} = d{\bf\alpha}[\phi,\delta_g\phi],
\label{sc3}
\ee
for some two-form ${\bf\alpha}$.  Combining \req{sc2} and \req{sc3}, we see
that $d{\bf j}[g] = 0$ when the equations of motion are satisfied, where
\be
{\bf j}[g]
  = {\bf\Theta}[\phi,\delta_g\phi] - {\bf\alpha}[\phi,\delta_g\phi] .
\label{sc4a}
\ee

The two-form ${\bf j}[g]$ is the Hodge dual of the usual Noether current
associated with the symmetry generated by $g$; its integral over a Cauchy
surface $\cal C$ gives a conserved charge $q[g]$.  Note that when ${\bf j}$
is exact, that is, ${\bf j} = d{\bf Q}$, then the integral that gives
$q[g]$ reduces to an integral at $\partial{\cal C}$.  This is the case
when the symmetry group is the group of diffeomorphisms, and explains why
the energy in general relativity can be written as an integral at spatial
infinity.

We now specialize to the case of diffeomorphism invariance.  Let
$\delta_\zeta\phi={\cal L}_\zeta\phi$ be a diffeomorphism generated by a
vector field $\zeta$.  It is then easy to show that
\be
{\bf j}[\zeta] = {\bf\Theta}[\phi,{\cal L}_\zeta\phi] - \zeta\cdot {\bf L} .
\label{sc6}
\ee
($\cal L$ is the Lie derivative, and the centered dot denotes contraction
of a vector with the first index of a form.)  The Noether charges now have
simple physical interpretations: if $t^\mu$ generates an asymptotic time
translation and $\varphi^\mu$ generates an asymptotic rotation, then Wald
has shown \cite{wald} that the canonical energy and angular momentum are
\bea
{\cal E} &=& \int_\infty ({\bf Q}[t] - t\cdot G)\nonumber\\
{\cal J} &=& -\int_\infty {\bf Q}[\varphi] .
\label{sc7}
\eea
Here the integrals are taken along a circle of constant time and infinite
radius, and $G$ is defined by the condition
\be
\delta_0 \int_\infty t\cdot G
  = \int_\infty t\cdot{\bf\Theta}[\phi,\delta_0\phi],
\label{sc8}
\ee
for variations $\delta_0$ lying within the space of solutions of the
equations of motion.

\medskip\noindent
{\bf B.  Noether Charges for Einstein Gravity}
\par
For the first-order Einstein action of \req{eq9}, it is easy to see that
\be
{\bf\Theta} = \half E^a\wedge\delta\omega_a ,
\label{newa}
\ee
and a simple computation gives a current (\ref{sc6}) of
\be
{\bf j}[\zeta] = d{\bf Q}[\zeta] , \qquad
{\bf Q}[\zeta] = -\half E^a\, \zeta\cdot\omega_a .
\label{eq13}
\ee
The canonical energy (\ref{sc7}) for the black hole solution of
\reqs{btztriad}--(\ref{btzconn}) is then easy to calculate: we obtain
\be
{\cal E} = \pi M ,
\label{eq15}
\ee
while the canonical angular momentum ${\cal J}$ is
\be
{\cal J} = \pi J  .
\label{eq17}
\ee

\medskip\noindent
{\bf C.  Noether Charges for the BCEA System}
\par
For the BCEA system, there are two sets of symmetries that may give rise
to interesting Noether charges, the diffeomorphisms and the I(ISO(2,1))
transformations (\ref{eq:11}).  These symmetries are not independent.
On shell,
\bea
{\cal L}_\zeta A^a &=& D(\zeta\cdot A^a),\nonumber\\
{\cal L}_\zeta B^a &=& D(\zeta\cdot B^a)
  + {1\over2}\epsilon^{abc}B_b (\zeta\cdot A_c),\nonumber\\
{\cal L}_\zeta C^a &=& D(\zeta\cdot C^a)
  + {1\over2}\epsilon^{abc}C_b (\zeta\cdot A_c),\nonumber\\
{\cal L}_\zeta E^a &=& D(\zeta\cdot E^a)
  + {1\over2}\epsilon^{abc}[E_b\zeta(\cdot A_c) + C_b(\zeta\cdot B_c)
  + B_b(\zeta\cdot C_c)] ,
\label{sc9b}
\eea
and the transformations (\ref{sc9b}) are precisely the gauge transformations
(\ref{eq:11}) with parameters
\be
\tau^a = \zeta\cdot A^a,\quad \rho^a = \zeta\cdot B^a,\quad
  \lambda^a = \zeta\cdot C^a,\quad {\hat\zeta}^a = \zeta\cdot E^a .
\label{sc10}
\ee
The diffeomorphisms are thus equivalent to I(ISO(2,1)) gauge transformations
on shell, as one expects in a topological theory.

{}From the Lagrangian (\ref{bceaaction}), it is evident that
\be
{\bf\Theta} = \half \left( E^a\wedge\delta A_a + B^a\wedge\delta C_a\right) .
\label{sc11}
\ee
For the transformations (\ref{eq:11}), it then follows that
\be
{\bf j}[g] = d{\bf Q}[g] , \qquad
{\bf Q}[g] = -\half\left( E^a\tau_a + B^a\lambda_a\right),
\ee
on shell.  The conserved charge is thus
\be
q[g] = \int_{\cal C}{\bf\Theta}[g]
     = -\half\int_\infty \left(E^a\tau_a + B^a\lambda_a \right).
\label{sc14}
\ee

In particular, consider the BCEA black hole.  For the integral (\ref{sc14})
to exist, we must restrict ourselves to parameters with the asymptotic
behavior
\be
\tau_a \sim {1\over r}\hat\tau_a , \qquad \lambda_a \sim \hat\lambda_a ,
\label{sc15}
\ee
where $\hat\tau_a$ and $\hat\lambda_a$ are constants (or possibly functions
of $\phi$).  Then
\be
q[\hat\tau_a,\hat\lambda_a]
  = {-\pi(-\hat\tau_0 r_- + \hat\tau_2 r_+)\over(r_+^2-r_-^2)^{1/2}}
  + \pi(\hat\lambda_0 r_+ + \hat\lambda_2 r_-) .
\label{sc16}
\ee
The charges are thus determined by the constants $r_\pm$, as one
might expect.

As a special case, we may consider transformations parametrized as in
(\ref{sc10}), that is, I(ISO(2,1)) transformations that correspond
to diffeomorphisms.  We find conserved charges
\bea
q[\zeta^\phi,\zeta^t] &=& - {\pi(r_+^2+r_-^2)\over\ell}\zeta^\phi
  + {2\pi r_+r_-\over\ell^2}\zeta^t \nonumber\\
  &=& -\pi M\ell\zeta^\phi + {\pi J\over\ell}\zeta^t .
\label{sc17}
\eea
To determine the canonical energy, we must add in the term $t\cdot G$ of
\req{sc7}, but it is not hard to check that this gives no further
contribution.  The charges associated with asymptotic time translations
and rotations are the mass and angular momentum, as expected, but
in the wrong order---the mass appears as the charge for $\zeta^\phi$, while
the angular momentum is the charge for $\zeta^t$!

A possible interpretation of this result is to attribute the mass and
charge of \req{sc17} to a combination of the black hole and the ``matter''
fields $B$ and $C$.  That is, we can write
\bea
M_{total} &=& M_{bh} + M_{B-C}\nonumber\\
J_{total} &=& J_{bh} + J_{B-C},
\label{sc18}
\eea
with
\be
M_{B-C} = -{J_{B-C}\over\ell} = {J\over\ell} - M .
\label{sc19}
\ee
It is perhaps unsurprising to find that the $B$ and $C$ fields carry angular
momentum---they are, after all, sources of torsion in the field equations
(\ref{eq:7d}).  We do not, however, have a good explanation for the extremal
condition $M_{B-C} = -J_{B-C}/\ell$.

\medskip\noindent
{\bf D.  Thermodynamics}
\par
In the first item of Ref.\ \cite{wald}, Wald shows that black hole
entropy can also be derived as a Noether charge.  Unfortunately,
that derivation breaks down in the first-order formulation.\footnote{In
particular, the ansatz used to define Wald's $\tilde{\bf Q}$ fails.}
However, we can still use the first law of thermodynamics to determine
an entropy for the BCEA system.

To do so, we choose a constant $\Omega_H$ such that the Killing vector
\be
\chi^\mu = t^\mu + \Omega_H \varphi^\mu,
\label{sce1}
\ee
vanishes at the horizon.  The surface gravity $\kappa$ is then
\be
\kappa^2 = -\half \nabla^\mu\chi^\nu\nabla_\mu\chi_\nu ,
\label{sce2}
\ee
and the first law of black hole dynamics takes the form
\be
{\kappa\over2\pi}\delta S = \delta{\cal E} - \Omega_H \delta{\cal J} .
\label{sce3}
\ee
For the (2+1)-dimensional black hole (\ref{eq5}), in particular,
it is easily checked that
\be
\Omega_H = -N^\phi(r_+), \qquad
\kappa = {r_+^2 - r_-^2\over r_+\ell^2} .
\label{sce4}
\ee
The relationship to standard thermodynamics may then be established
by means of the Euclidean path integral: it is shown in Refs.\
\cite{btz,Euclbh} that the black hole temperature obtained from the
path integral is $\beta=2\pi/\kappa$, so (\ref{sce3}) is just the
first law of thermodynamics.

For the black hole in Einstein gravity, (\ref{eq15})--(\ref{eq17})
gives
\be
\delta{\cal E} - \Omega_H \delta{\cal J} = {2\pi\over\ell^2}
  \left\{ r_+\delta r_+ + r_-\delta r_-
  - {r_-\over r_+}(r_+\delta r_- + r_-\delta r_+)\right\}
= 2\pi\kappa\delta r_+ .
\label{sce5}
\ee
The entropy (\ref{sce3}) is thus
\be
S = 4\pi^2 r_+ ,
\label{sce6}
\ee
agreeing with the conventional expression in our choice of units.
For the BCEA black hole, on the other hand, (\ref{sc17}) gives
\be
\delta{\cal E} - \Omega_H \delta{\cal J} = {2\pi\over\ell^2}
  \left\{ r_+\delta r_- + r_-\delta r_+
  - {r_-\over r_+}(r_+\delta r_+ + r_-\delta r_-)\right\}
= 2\pi\kappa\delta r_- .
\label{sce7}
\ee
Hence
\be
S = 4\pi^2 r_- .
\label{sce8}
\ee
Once again, the parameters of the conventional black hole have been
interchanged: this time, the roles of $r_+$ and $r_-$ are reversed.

\bigskip
\section{Conclusion}
We have seen that the (2+1)-dimensional black hole does not require a
negative cosmological constant.  Rather, the cosmological constant can be
replaced by a suitable distribution of ``matter.''  In itself, this is
perhaps not a very surprising observation, although the division of mass
and angular momentum between the metric and the $B$ and $C$ fields is
rather unexpected.  But it is certainly interesting that even such a simple
system of topological matter can give rise to a black hole.

One may also examine the BCEA model in second order form, using (\ref{aa0})
and (\ref{aa}) to express the connection $A^a$ in terms of the fields
$E^a$, $B^a$, and $C^a$ in the action.  In this case, it turns out that the
$B$ and $C$ fields contribute to the Noether current at spatial infinity,
but not at the event horizon.  The result is that, as in the first order
form, the locally constructed expression for the entropy does not satisfy
the first law of thermodynamics.  This is probably generic in models
where matter fields are coupled to the spin connection, {\it i.e.}, to
first derivatives of the metric.

The BCEA system is, in fact, quite powerful, and our results suggest an
interesting direction to search for additional black hole solutions.
As noted in section 1, the standard (2+1)-dimensional black hole can
be obtained from anti-de Sitter space by a set of identifications.  It is
natural to ask whether this procedure can lead to black hole solutions if
we begin with a different homogeneous geometry.

Homogeneous Riemannian metrics have been classified by Thurston
\cite{thurston78}, and their Wick rotated metrics provide a useful
starting point.  Our preliminary results are that ``black-hole-like''
solutions can be obtained from at least three of these geometries,
$H^3$ (which gives the usual black hole), $H^2\times E^1$, and {\it SOL}.
In particular, if we let
\be
f(r):={r^2\over \ell^2}-M ,
\label{eq:f}
\ee
the metrics
\be
ds^2=-f(r)dt^2+{1\over f(r)}dr^2 +d\phi^2,
\ee
(obtained from $H^2\times E^1$ by identifications) and
\be
ds^2=
 -{f\over M\ell^2}dt^2+{1\over\ell^2 f}dr^2 +{M\over f}e^{-2t/\ell}d\phi^2.
\ee
(obtained from {\it SOL}) satisfy the BCEA equations of motion for suitable
choices of $B$ and $C$.  We do not yet understand the detailed behavior of
these solutions, but we believe them to be of some interest.

\bigskip\noindent
{\bf ACKNOWLEDGEMENTS}

\noindent
One of the authors (SC) was supported in part by National
Science Foundation grant
PHY-93-57203 and Department of Energy grant DE-FG03-91ER40674.
Two of the authors (JG and RBM) would like to acknowledge the partial support
of the Natural Sciences and Engineering Research Council of Canada.

\end{document}